\pdfoutput=1 
\documentclass{JINST}
\usepackage{upgreek}
\usepackage{subfigure}
\usepackage{epstopdf}
\usepackage[percent]{overpic}

\title{Application of the TPB Wavelength Shifter to the DEAP-3600 Spherical Acrylic Vessel Inner Surface}

\author{B.~Broerman$^a$\thanks{Corresponding author.}~, M.\,G.~Boulay$^{a,b}$, B.~Cai$^a$, D.~Cranshaw$^a$, K.~Dering$^{a}$, S.~Florian$^a$, R.~Gagnon$^{a}$, P.~Giampa$^a$, C.~Gilmour$^a$, C.~Hearns$^{a}$, J.~Kezwer$^a$, M.~Ku{\'z}niak$^a$\thanks{Current address: Department of Physics, Carleton University, Ottawa, Ontario K1S 5B6, Canada}, T.~Pollmann$^c$\thanks{Current address: Technische Universit\"at M\"unchen, 80333 Munich, Germany}, M.~Ward$^a$\thanks{Current address: Department of Physics, Royal Holloway, University of London, Egham Hill, Egham, Surrey TW20 0EX, United Kingdom}\\
\llap{$^a$}Department of Physics, Engineering Physics and Astronomy, Queen's University,\\
  Kingston, Ontario K7L 3N6, Canada\\
\llap{$^b$}Department of Physics, Carleton University,\\
  Ottawa, Ontario K1S 5B6, Canada\\
\llap{$^c$}SNOLAB, Laurentian University,\\
  Sudbury, Ontario P3E 2C6, Canada\\

E-mail: \email{broerman@owl.phy.queensu.ca}}

\abstract{DEAP-3600 uses liquid argon contained in a spherical acrylic vessel as a target medium to perform a sensitive spin-independent dark matter search. Argon scintillates in the vacuum ultraviolet spectrum, which requires wavelength shifting to convert the VUV photons to visible so they can be transmitted through the acrylic light guides and detected by the surrounding photomultiplier tubes. The wavelength shifter 1,1,4,4-tetraphenyl-1,3-butadiene was evaporatively deposited to the inner surface of the acrylic vessel under vacuum. Two evaporations were performed on the DEAP-3600 acrylic vessel with an estimated coating thickness of 3.00~$\pm$~0.02~$\upmu$m which is successfully wavelength shifting with liquid argon in the detector. Details on the wavelength shifter coating requirements, deposition source, testing, and final performance are presented.}


\begin{document}

\section{Introduction}
\label{sec:Intro}
\begin{figure}
\begin{center}
	\includegraphics[width=4in]{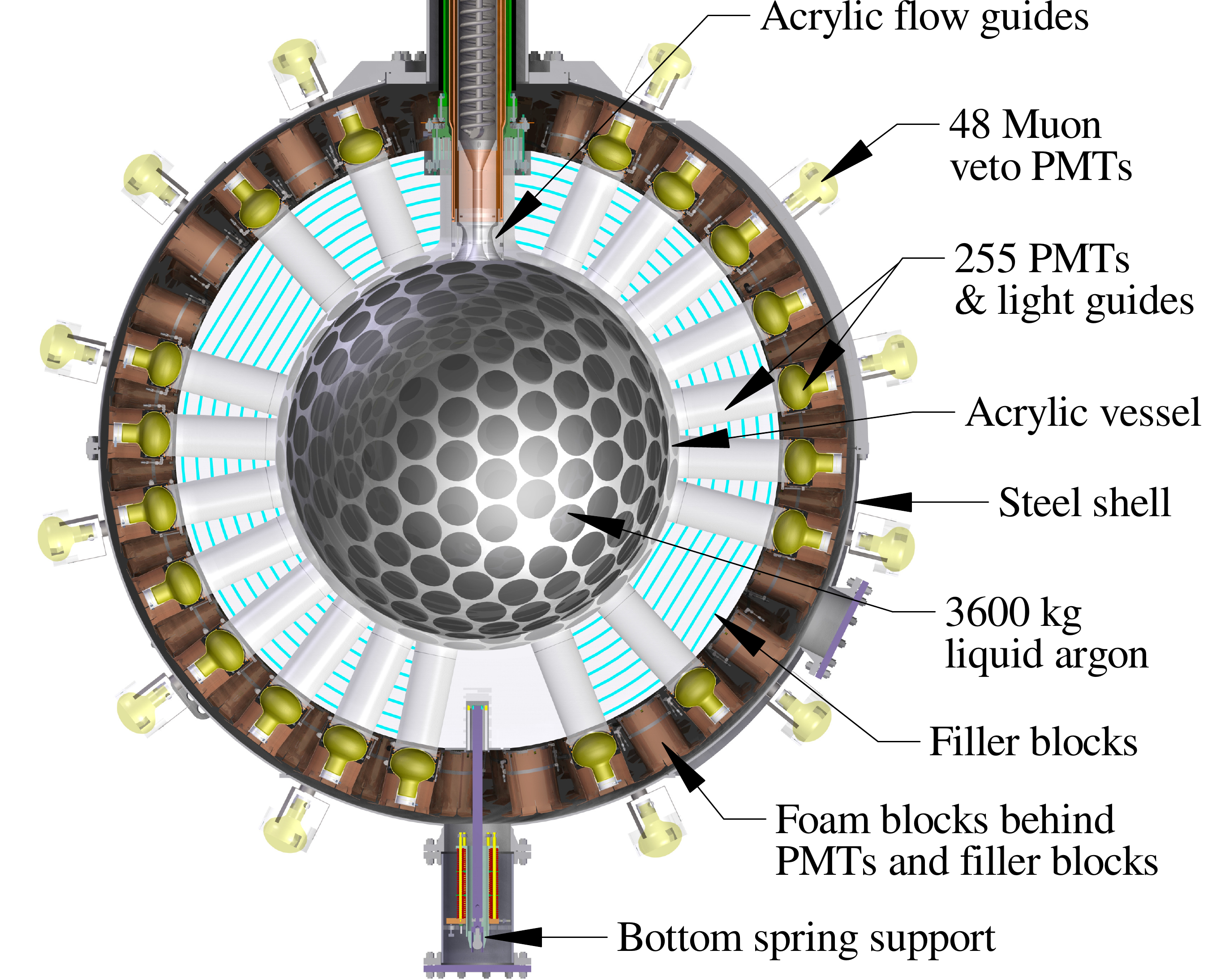}
\caption{Schematic of the DEAP-3600 detector. 255~inward-facing PMTs view the inner spherical cavity containing the LAr. A coating of the wavelength shifter 1,1,4,4-tetraphenyl-1,3-butadiene was applied over the inner surface of the acrylic vessel (not annotated) prior to the insertion of the acrylic flow guides and filling with LAr.}
\label{fig:deap}
\end{center}
\end{figure}

Observational evidence supports a model of cosmology, $\Lambda$CDM, which proportions approximately 85$\%$ of the mass of the universe to a yet-unidentified constituent: dark matter~\cite{planck2015planck}. A leading dark matter candidate, the Weakly Interacting Massive Particle (WIMP), lies outside the current standard model of particle physics, though may be detectable through its scattering from nuclei in a laboratory-based experiment. DEAP-3600, located 2~km underground at SNOLAB in Sudbury, Canada, is a single phase liquid argon dark matter search experiment~\cite{amaudruz2014deap}, shown schematically in Figure~\ref{fig:deap}. The liquid argon (LAr) is contained in a spherical acrylic vessel (AV) 85~cm in radius and viewed by 255~inward-facing photomultiplier tubes (PMTs) through 50~cm acrylic light guides. With a 1000-kg fiducial mass, it aims to perform a spin-independent dark matter search targeted at a WIMP-nucleon cross section of 10${}^{-46}$~cm$^{2}$ for a 100~GeV$/c^2$ WIMP mass. 

Argon scintillates in the vacuum ultraviolet (VUV) spectrum from decaying dimer states resultant of particle interactions. As VUV photons are efficiently absorbed by most optical media (acrylic, PMT glass), information from the deposition of energy in the LAr volume must be wavelength shifted before it can be recorded by the surrounding PMTs. The organic wavelength shifter, 1,1,4,4-tetraphenyl-1,3-butadiene (TPB), evaporatively deposited over the approximately 9-m$^2$ inner surface of the AV prior to filling with LAr, absorbs the VUV light and re-emits in the visible spectrum, peaked near 420~nm~\cite{gehman2011fluorescence}. 

TPB is well-suited for liquid argon rare-event searches and long-term stability of coatings has been demonstrated experimentally~\cite{agnes2015first}. It has a fast singlet re-emission time on the order of 1~ns~\cite{flournoy1994substituted} and a high conversion efficiency for VUV light aligning with the region of high PMT quantum efficiency. Additionally, it is a highly refined material, the production of which can be carefully controlled to limit radio-contaminant exposure. 

The requirements for the TPB coating will be discussed in Section~\ref{sec:requirements}, design and operation of the evaporation source in Section~\ref{sec:design_operation}, construction of the source in Section~\ref{sec:construction}, test evaporations in Section~\ref{sec:testing}, and the final deposition on the DEAP-3600 detector vessel and ex situ analysis of the coating in Section~\ref{sec:final_deposition}.

\section{Coating requirements}
\label{sec:requirements}
The acrylic for the AV was exposed to natural air during construction, allowing $^{222}$Rn, with a diffusion length in room temperature acrylic of approximately 0.1~mm~\cite{wojcik1991measurement}, to diffuse into the acrylic. To reach the internal background target of less than 1~event in 3~years running, the inner surface of the AV was sanded in situ with a robotic resurfacer to remove a significant portion of the absorbed and diffused-in radio-contaminants. As the clean AV was then exposed to the TPB deposition system, a strict radon emanation requirement was set for all components of the source and deployment system of 10~mBq, to prevent the reintroduction of radon and radon progeny on the inner AV surface. 

To address purity of the TPB, special considerations by the manufacturer (Alfa Aesar; Heysham, England) provided TPB that was produced from only certified high-purity reagents and with the synthesis process occurring under N$_{2}$ cover gas to prevent exposure to radon-laden air. 

\begin{figure}
\begin{center}
\includegraphics[width=3.5in]{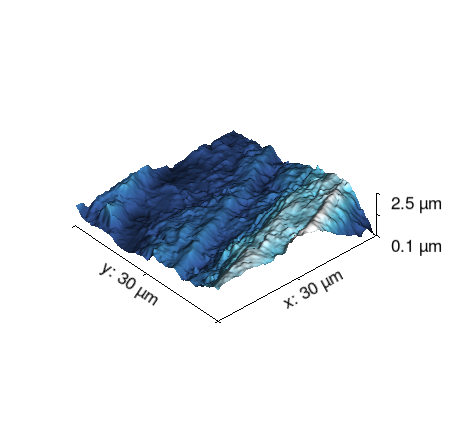}
\includegraphics[width=3.5in]{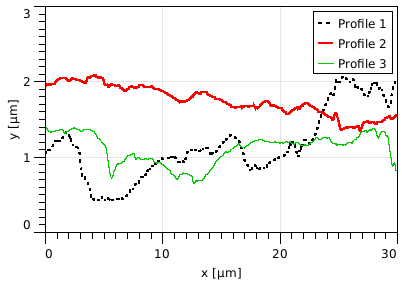}
\caption{AFM scan of an acrylic test sample machine-sanded with the same sandpaper used for resurfacing the DEAP-3600 acrylic vessel. Top panel: a representative 2D profile 30 by 30~microns in size. Bottom panel shows three example 1D profiles from the same sample: (1) across and (2) along the sanding marks, (3) along the X-axis. Features of approximately 1-2~$\upmu$m height are present.}
\label{fig:afm}
\end{center}
\end{figure}

The roughness features on the inner acrylic surface left from the resurfacer sanding paper span thicknesses of 1-2~$\upmu$m, as shown in Figure~\ref{fig:afm} from atomic force microscopy (AFM) scans of sanded test acrylic. This surface roughness sets the targeted thickness for the applied TPB coating; the TPB coating must be at least as thick as the variation in the underlying surface morphology. Complete coverage of the inner surface of the AV with the wavelength shifter was necessitated to ensure the collection of as much of the argon scintillation light as possible to reach the targeted 8~PE/keV detector light yield. The target thickness was hence fixed to 3~$\upmu$m. Conversion efficiency of the coating does not depend on the thickness in $\sim$2-6~$\upmu$m range, as demonstrated by other groups~\cite{Francini}.

Across large-scale areas of the inner surface, the TPB coating was specified not to exceed a thickness non-uniformity of 20\% which could lead to biases in light yield and position reconstruction across the detector. Although very little change of conversion efficiency with coating thickness is expected, increased visible light scattering expected from a thicker coating could result in asymmetries in visible light propagation across the detector. It was required for the thickness variation to be smaller than the scattering length in TPB, which was assumed to be 1~$\upmu$m; the scattering length in TPB is weakly constrained with only one published measurement of 2-3~$\upmu$m~\cite{Stolp}, published recently.

Additionally, no TPB was to be applied to the inner surface of the detector neck, realized through the use of a baffle described in Section~\ref{sec:final_deposition}.

During the initial phase of the liquid fill in the DEAP-3600 detector, the bottom surface of the acrylic vessel was exposed to liquid argon dripping from the above cooling coil. This can potentially generate a large, localized thermal stress in the TPB coating. In order to test the robustness of the coating on acrylic surfaces to the exposure of cryogenic liquids, test acrylic slides were coated with 1~$\upmu$m of TPB and either dipped in liquid nitrogen, or had liquid nitrogen poured on them from a representative height of approximately 2~m. Some adverse effects, including flaking, were observed in the absence of controlled measures to keep the samples dry, due to humidity from the air freezing out on the samples. No flaking, nor coating degradation, was noticed provided the coatings and conditions of the test were kept dry --- the conditions maintained during the DEAP-3600 deposition.

\section{Source design and operation principle}
\label{sec:design_operation}

\begin{figure}[h!]
 \centering
  \begin{minipage}[b]{0.45\textwidth}~~~~~
    \includegraphics[width=0.85\textwidth]{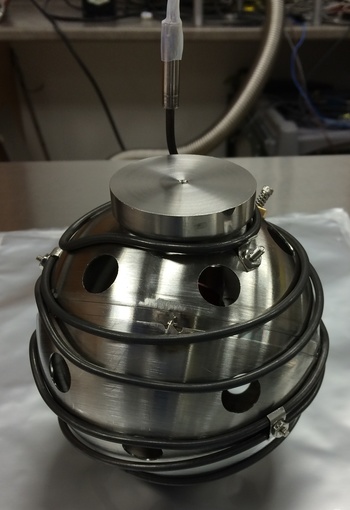}
   \caption{Final stainless steel TPB deposition source. The coil heater is held in thermal contact with the sphere with compressed tabs.}
   \label{fig:source}
  \end{minipage}
  \hspace{0.05\linewidth}
  \begin{minipage}[b]{0.45\textwidth}~~~~~~~~
    \includegraphics[width=0.70\textwidth]{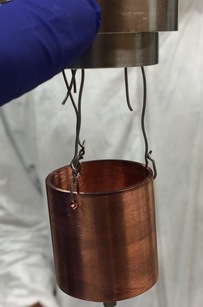}
    \caption{The inner copper crucible to hold the TPB powder is attached to a stainless steel top hat and lowered into the spherical source.}
    \label{fig:crucible}
 \end{minipage}
\end{figure}

TPB was deposited on the inside of the AV using a spherical evaporation source deployed to the centre of the AV. The source, shown in Figure~\ref{fig:source}, is an 11-cm-diameter hollow sphere with 20, 14-mm-diameter holes located at the vertices of an inscribed dodecahedron. 

Modeling of the geometric coverage from multiple deposition sources, each with an evaporation profile found in Reference~\cite{tinaThesis}, in addition to Monte Carlo simulation~\cite{kersevan2009introduction} showed 20~equally-spaced holes provided sufficient coverage and uniformity across the enveloping spherical acrylic substrate, while maintaining feasibility of construction. The holes were orientated to provide a solid face at the top to limit preferential upward evaporation during warm-up of the crucible, and a face at the bottom to prevent TPB powder falling during deployment.

A 34~mm by 35~mm cylindrical copper crucible, shown in Figure~\ref{fig:crucible} and into which the TPB powder is loaded, hangs in the centre of the hollow sphere and is radiatively heated with a Watlow coil heater externally wrapping the source. The external sphere is heated, rather than the crucible directly, to ensure the sphere is at the TPB sublimation temperature of approximately 208${}^{\circ}$C before the crucible, to promote desorption off the surface and proper performance of the source. Uniformity of the emitted flux from the source is accomplished by scattering TPB molecules off the inner surface of the evaporation sphere after leaving the crucible, which randomizes their motion while inside the source.

A controlled and quality coating is achieved by maintaining a slow deposition rate below 5~\AA/s. Additionally, a deposition rate of 1~\AA/s allows the entire DEAP-3600 deposition to occur in 1-2~underground shifts, simplifying monitoring and personnel.

Once the TPB molecules leave the source, to ensure no scattering along the 0.85-m distance to the inner surface of the AV, a vacuum better than approximately $1\times 10^{-5}$~mbar is desired. Ultra-high vacuum modeling of the deposition system, which included the source, AV, and vacuum system and assumed diffuse scattering inside the source, showed the pressure outside the TPB source falls to the background vacuum pressure within approximately 2~cm past the source radius. 
 
\section{Evaporation source construction}
\label{sec:construction}
A prototype source was constructed for testing and verifying the principle of operation. As radio-cleanliness was beyond the scope of these tests, the prototype sources were constructed of aluminum and heated with bare nickel-chrome (Nichrome) wire wrapped in Kapton insulation. 

The source for the DEAP-3600 deposition was machined of two 316~stainless steel hemispheres and tack-welded with an equatorial band to form a sphere. The Watlow 125CH93A1X coil heater, which consists of a Nichrome heating wire encapsulated in an electrically insulated stainless steel sheath, was attached with stainless steel clips after the holes were machined in the source. The inner copper crucible was bored from a solid copper rod.

After machining, all components were ultrasonically cleaned in both Alconox detergent and ultra-pure water (UPW) baths. A 10\% by weight citric acid bath was prepared for passivation of the source components. The source and crucible were then baked at 210$^{\circ}$C in a vacuum chamber for 5~days. After baking, the source and crucible were immediately bagged and heat-sealed in a multi-layer DuPont liner material composed of nylon, aluminum foil, and low density polyethylene layers, before being transferred underground to SNOLAB. 

Further details on the construction of the final source can be found in Reference~\cite{broermanThesis}.

\section{Test depositions}
\label{sec:testing}
\begin{figure}[h!]
  \centering
  \begin{minipage}[b]{0.4\textwidth}
    \includegraphics[width=\textwidth]{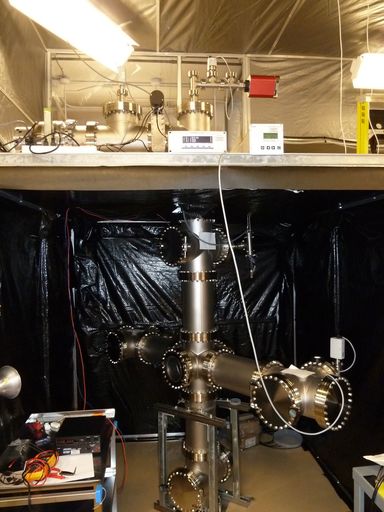}
    \caption{The initial vacuum system and dark box built for testing of the prototype TPB source. Much of the vacuum pumping hardware was used on the final DEAP-3600 deposition. Notable here, quick-access doors at the ends of the cross allowed for rapid access to glass samples.}
    \label{fig:vacuumCross}
  \end{minipage}
  \hspace{0.05\textwidth}
  \begin{minipage}[b]{0.4\textwidth}
    \includegraphics[width=\textwidth]{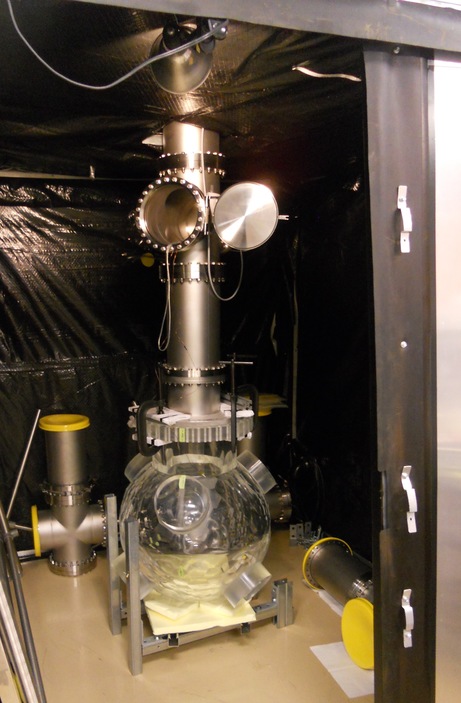}
    \caption{Rebuilt vacuum system with a 20-inch-diameter test acrylic vessel. The vacuum pumping hardware from the initial system was reused. The test vessel is sealed to the evacuation chamber with a custom o-ring flange adaptor.}
    \label{fig:vacShallow}
  \end{minipage}
\end{figure}

A discussion of initial testing with a prototype TPB evaporation source can be found in Reference~\cite{tinaThesis}. The vacuum system for these tests, shown in Figure~\ref{fig:vacuumCross}, was cross-shaped, with its four arms being of comparable length to the 85~cm DEAP-3600 AV inner radius. Glass and acrylic sample disks were positioned at varying distances from the source inside this cross-shaped vacuum chamber to evaluate source performance and investigate coating quality. With the addition of the small test evaporation system in~\cite{tinaThesis}, TPB coatings ranging in thickness from 0.5~$\upmu$m to 4~$\upmu$m were tested for application in the DEAP-1 prototype and DEAP-3600~\cite{Deap1Rn}.

\begin{figure}[h!]
\begin{center}
	\includegraphics[width=4.75in]{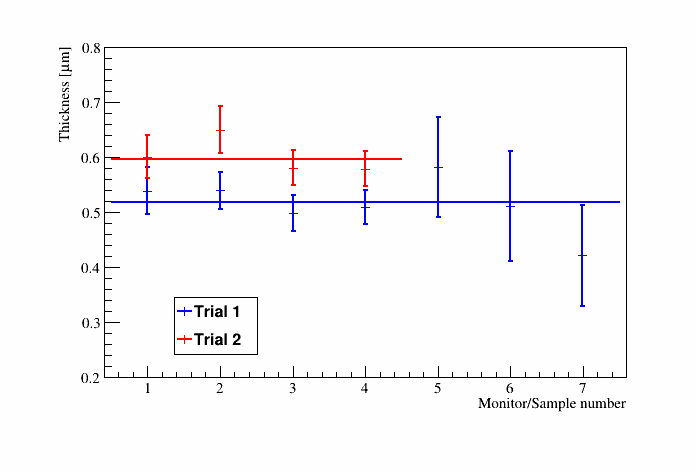}
	\caption{Distribution of TPB thicknesses at a distance of 85~cm from the source during TPB testing. Abscissae numbers 1--4~are the readings from the deposition monitors. Numbers 5--7~are glass and acrylic samples, measured with a stylus profiler. A fit to the data is shown in the solid lines.}
	\label{fig:tinaMeasuredThickness}
\end{center}
\end{figure}

Thickness data from two test depositions are shown in Figure~\ref{fig:tinaMeasuredThickness}, collected from Inficon SL-A0E40 front-loaded quartz deposition monitors (abscissae 1--4) and glass samples (abscissae 5--7) after measurement with a Dektak 8M stylus profiler. The quartz monitors measure the thickness of an evaporant deposited onto the crystal diaphragm through the downshift in resonant frequency of the crystal caused from the increase in deposited mass. The larger uncertainty in thickness from the glass samples arises from the resolution of the stylus profiler used to measure the coating. To account for the varying distances of the targets, the measured thickness was extrapolated to that expected from being 85~cm from the source, based on the ratio of solid angles. Trials 1~and 2~had 4.2~g and 6.3~g TPB loaded into the crucible, respectively, with an expected thicknesses of 0.43~$\upmu$m and 0.60~$\upmu$m, respectively, at an 85~cm sampling distance. The expected thickness for Trial 1 is nominally greater than the expected thickness, as the readings were taken while the deposition monitors had not yet cooled to ambient temperature from being warmed by the TPB source. This cooling reduces the thickness reading from the quartz monitors.

\begin{figure}[h!]
\begin{center}
	\includegraphics[width=4in]{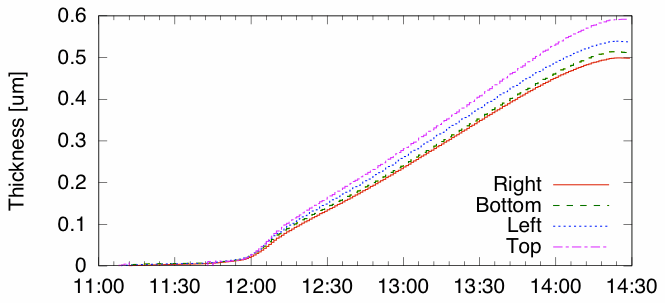}
	\caption{Deposition monitor data during testing from four locations in the vacuum system at a distance approximately 85~cm from the source. The 0.1~$\upmu$m difference between the right and top sensors shows non-uniformity to less than 20\%. }
	\label{fig:depCurves}
\end{center}
\end{figure}

Figure~\ref{fig:depCurves} shows the deposited thickness as a function of time during one evaporation for the four deposition monitors located in the arms of the system approximately 85~cm from the source. After initial warm-up, the profile follows a stable deposition rate followed by a gradual tailing-off as the TPB is emptied from the crucible. Agreement between the four locations implying a measured non-uniformity in thickness below 20\% provided justification for a full scale test.  

\begin{figure}[h!]
\begin{center}
	\includegraphics[width=3.7in]{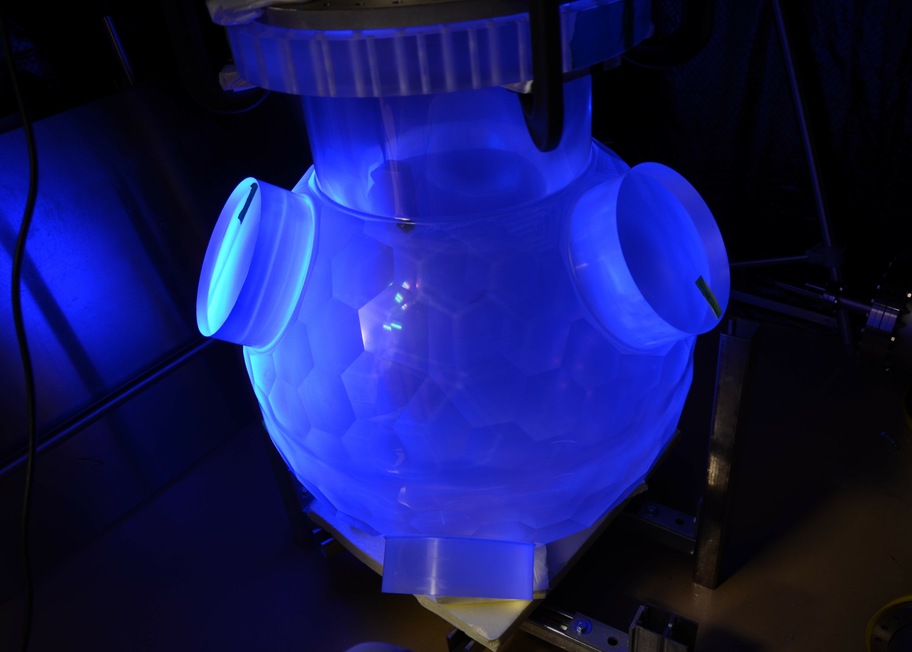}
\caption{The 20-inch-diameter test vessel after the first TPB deposition and under excitation with a UV lamp. Full coverage of the inner surface was achieved.}
\label{Shallow}
\end{center}
\end{figure}

The first large-scale 4$\pi$ deposition was performed in April 2013. A 20-inch-diameter acrylic test vessel was fitted to the vacuum system, shown in Figure~\ref{fig:vacShallow}. In this test, which aimed at verifying the ability to coat an inclosed spherical surface, TPB was evaporated to the unsanded inner surface to a thickness of approximately 1~$\upmu$m as read by a quartz deposition monitor placed near the radius of the test vessel's inner surface. Figure~\ref{Shallow} shows the coated test vessel, backlit with a 265~nm ultraviolet lamp for qualitative examination of the coating. No large-scale features or defects in the coating, nor areas uncoated, were noticed. 

\section{Evaporation on the DEAP-3600 detector vessel}
\label{sec:final_deposition}

\subsection{Deployment system}
\begin{figure}[h!]
\begin{center}
	\includegraphics[width=5in]{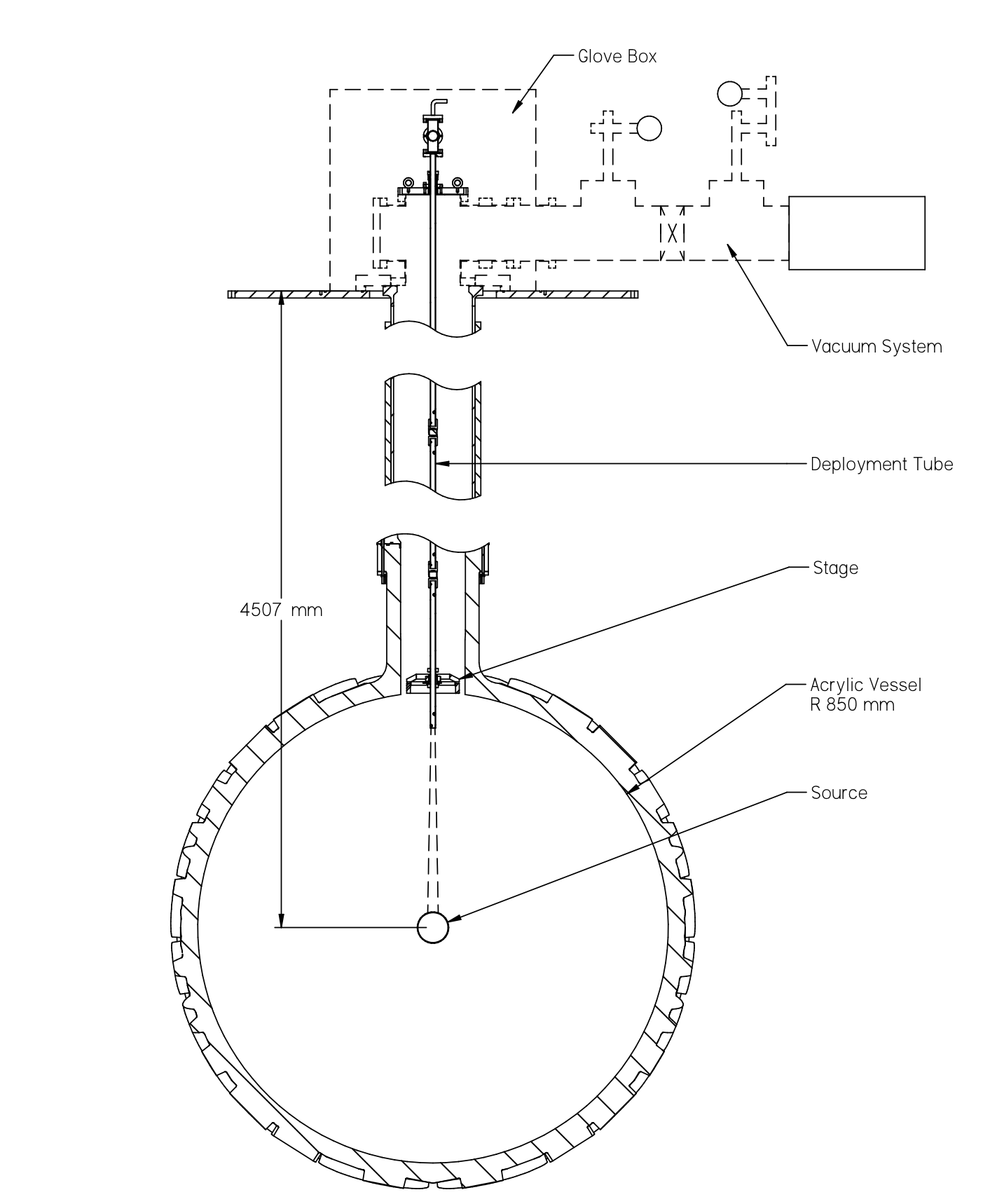}
\caption{Deployment tube for source access inside the AV. A 4-m long articulated tube traversed the detector neck with 10-inch~ConFlat sealing top flange inside the N$_{2}$ purged glovebox. The source is attached to the end of the tube with stainless steel cables. A stage, located at the intersection of the sphere and neck, acted as a bumper, baffle, and mount for the deposition monitor and sample slides. Not shown schematically are the filler blocks and PMTs surrounding the AV.}
\label{fig:DeploymentSystem}
\end{center}
\end{figure}

The source was lowered to the centre of the DEAP-3600 AV at the end of a four meter 0.75-inch-diameter stainless steel tube articulated in two locations with stainless steel universal joints, shown schematically in Figure~\ref{fig:DeploymentSystem}. A top sealing flange on the deployment tube mated to a custom 5-way cross joining the AV-neck space to the vacuum pump out branch, consisting of high vacuum-rated components. Driven by a 1250-l/s Pfeiffer HiPace turbomolecular pump (TMP), the vacuum branch included high and low pressure gauges, as well as purge inlets and outlets on both sides of a 10-inch ConFlat GTMP-8002 NorCal gatevalve which was hardware-interlocked to shut in the event of a power outage or pressure spike. This would isolate the AV to prevent lab air back-streaming through the pump in the event of a power outage, and protect the TMP from atmospheric exposure at full rotation speed in the event of an upstream leak. 

Two F3105 Thin Film Resistance Temperature Detectors (RTDs, OMEGA Engineering Inc.), one on the source surface and one on the inner copper crucible, were used to monitor the temperature of the source and were read out with an industrial DeltaV slow control system. The Watlow heating element was controlled from a power relay (Watlow DIN-A-MITE~A) which varied the duty cycle of a 120-V line-level supply. Software interlocks on source temperature and AV vacuum conditions linked a control signal to this main relay. In the event of source over-heating or a fault in the vacuum system, power to the heater would be shut down.

Two PTFE-insulated voltage cables, two 4-wire Kapton-insulated Lakeshore Cryogenic cables for RTD readout, and one Temp-flex 50HCX-18 coaxial cable for deposition monitor readout were routed through the deployment tube and out of the vacuum space through a 2.75~inch ConFlat 4-way cross located at the top of the tube. Additionally, a fluorinated ethylene propylene (FEP) purge line was run to let the AV back to lab pressure from vacuum with radon-scrubbed boil-off nitrogen. All source components were selected for adherence to the radiopurity requirement of an ultra-low radon emanation rate, high temperature resistance, and vacuum compatibility.

An acrylic stage (shown from below in Figure~\ref{fig:understage}) located at the intersection of the AV sphere and neck, was mounted with a quartz deposition monitor described in Section~\ref{sec:testing} to record and control the real-time thickness of the applied coating. The stage was further skirted with Tyvek sheeting to prevent extraneous TPB coating on the neck, while allowing conduction for the vacuum required for the deposition. Three acrylic disks, two of which were sanded with the sandpaper used by the resurfacer to a similar surface roughness, acted as witness plates. They were mounted to the under side of the stage in view of the TPB source to sample the coating for ex situ surface analysis. 

\begin{figure}[h!]
\begin{center}
	\includegraphics[width=3in]{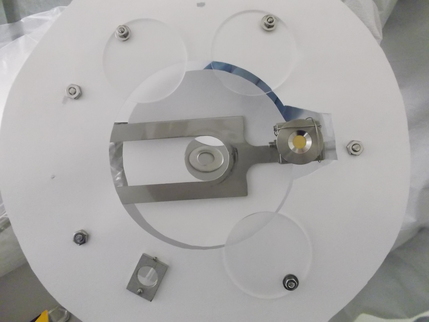}
\caption{Underside of the stage deployed at the sphere-neck junction. Tyvek skirting prevented extraneous TPB coating up the neck. Also shown are the three acrylic witness sample slides and the deposition monitor.}
\label{fig:understage}
\end{center}
\end{figure}

\subsection{Procedure and deposition}
Prior to deposition of the wavelength shifter, the inner surface of the DEAP-3600 acrylic vessel was prepared with a vacuum bake in an effort to remove absorbed water from the post-resurfacing UPW rinses. Using the empty TPB source as a heating element, the inner surface of the acrylic was raised to approximately 50${}^{\circ}$C while remaining under vacuum. This addressed both the overall gas load allowing for a lower attainable ultimate pressure in the AV, and removed water which may quench the argon scintillation signal~\cite{thonnard1972time}. In the end, the gas load from the AV was dominated by water diffusion through the bulk acrylic shell. 

Owing to the considerations on the thickness of the coating, and the limited packing space inside the crucible, two layers of TPB were applied. In each deposition, after loading of the TPB, the source was placed in the glovebox and brought to 150$^{\circ}$C under a rough vacuum for a pre-bake of the TPB to remove low-evaporation point impurities. The source was then allowed to cool before being deployed into the AV. After being pumped to a base pressure of approximately $1\times 10^{-5}$~mbar\footnote{The inner AV pressure was extrapolated from vacuum gauges higher in the detector neck using the ultra-high vacuum simulation package MolFlow~\cite{kersevan2009introduction}.} to ensure proper performance of the system, the temperature of the source was gradually raised until a stable deposition rate of 0.7~\AA/s, as viewed in real-time by the quartz deposition monitor in the neck, was achieved. After the first deposition, the AV volume and TPB surface were kept under vacuum or radon-scrubbed boil-off N$_{2}$ purge gas to remove the risk of a layer of surface impurities forming before the second TPB layer was applied. 

A fully accurate mapping of the TPB thickness across the approximately 9~$\rm{m}^2$ inner AV surface is impossible to produce. Nevertheless, an estimation of the average coating thickness can be made under the assumption that the source operated as expected to produce a uniform deposition flux. Based on the total mass loaded into the crucible of 29.4~$\pm$~0.2~g, the expected uniform coating would have a thickness of 3.00~$\pm$~0.02~$\upmu$m.

\subsection{Ex situ analysis of the coating}
After extraction of the TPB system from the AV, the sample acrylic slides attached to the under side of the stage were removed. Under scanning electron (FE-SEM Zeiss Gemini) and optical microscopes (Nikon Eclipse Ni-U), TPB coverage of the sample acrylic substrate was confirmed.

\begin{figure}[h!]
\centering
\subfigure[Profile SEM]{
	\includegraphics[width=4in]{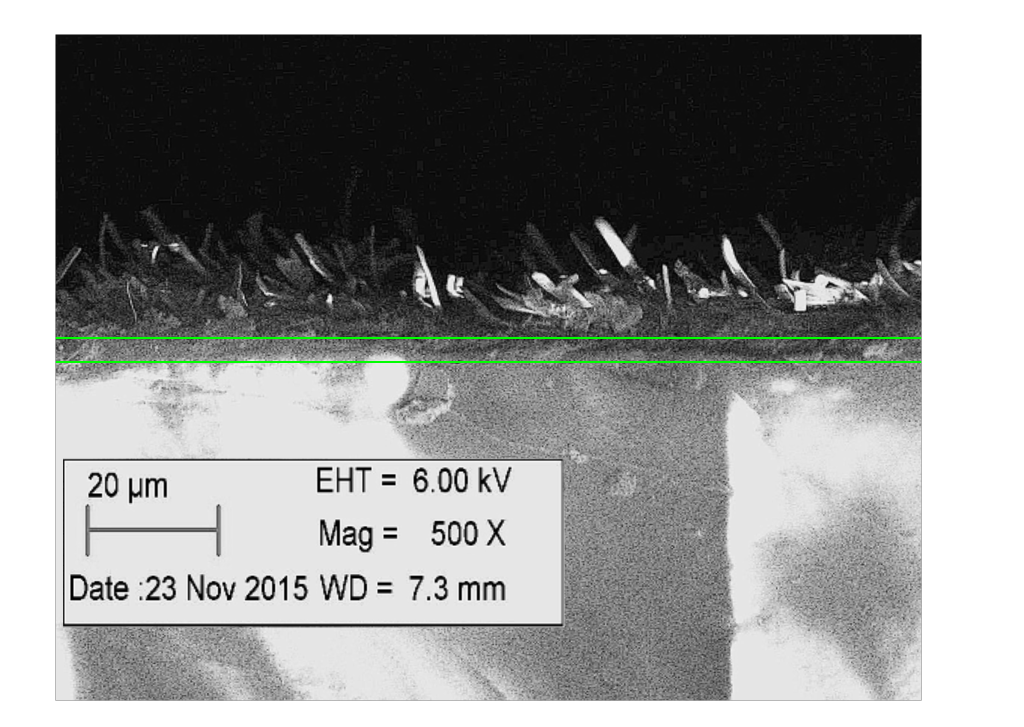}
	\label{fig:SEMprofileCursor}}
\subfigure[Aerial SEM]{
	\includegraphics[width=4.5in]{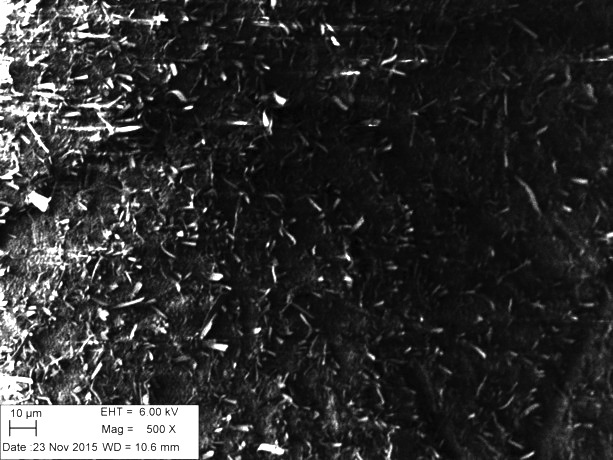}
	\label{fig:SEMprofile}}
\caption{SEM images of acrylic witness samples after deposition and extraction from the vessel. a) Profile scan of a cleaved sample edge with a cursor set at 3~$\upmu$m. Visible here are vertical needle-like $\alpha$-TPB polymorphic structures b) Aerial scan showing the distribution of $\alpha$-TPB structures.}
\label{fig:SEM}
\end{figure}

Although TPB is an insulator, an additional metallic coating, typically used in such cases to improve contrast and resolution in the SEM, was not applied to the studied surface. This allowed for the preservation of the TPB surface features, at the price of reduced image clarity. Profile SEM scans of cleaved edges of the sample slides corroborated, qualitatively,  a base-layer of TPB with a thickness of approximately 3~$\upmu$m and importantly the coverage of the underlying acrylic features, shown in Figure~\ref{fig:SEMprofileCursor} with a cursor set at 3~$\upmu$m. 

\begin{figure}[h!]
\begin{center}
	\begin{overpic}[width=4.5in]{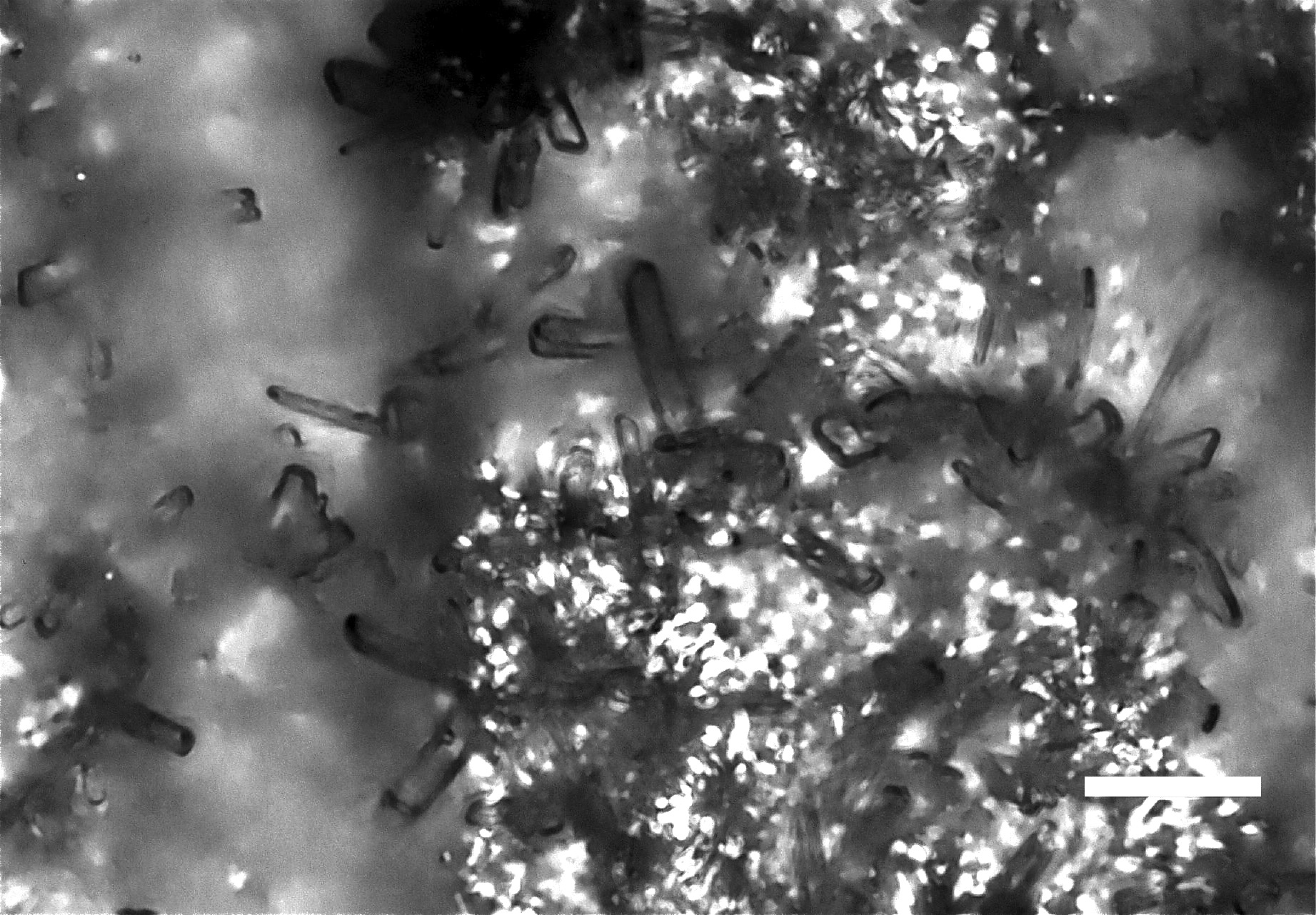}
        \linethickness{1pt}\put(84.1,9.3){\line(1, 0){10.0}}
        \linethickness{1pt}\put(84.1,9.3){\line(0, 1){1} }
        \linethickness{1pt}\put(94.1,9.3){\line(0, 1){1} }
        \end{overpic}
	\caption{Aerial optical microscope image of the coating showing needle-like $\alpha$-TPB structures. The scale bar corresponds to 6.25~$\upmu$m. This image was produced with a focus stacking technique by combining multiple photographs taken at different focal depths.}
	\label{fig:layeredOptical}
\end{center}
\end{figure}

Revealed in Figure~\ref{fig:SEMprofileCursor}, and additionally in aerial SEM and layered optical scans, Figures~\ref{fig:SEMprofile} and \ref{fig:layeredOptical}, respectively, are crystallized needle-like TPB structures extending vertically above the base layer. These structures ranged up to a height of approximately 10~$\upmu$m, and are $\alpha$-TPB, the most commonly encountered polymorph~\cite{bacchi2013raman}. The growth of $\alpha$-TPB is a result of sublimation followed by slow cooling and crystallization, which is the deposition temperature profile expected from the DEAP-3600 depositions.

\begin{figure}[h!]
\begin{center}
	\includegraphics[width=0.45\textwidth]{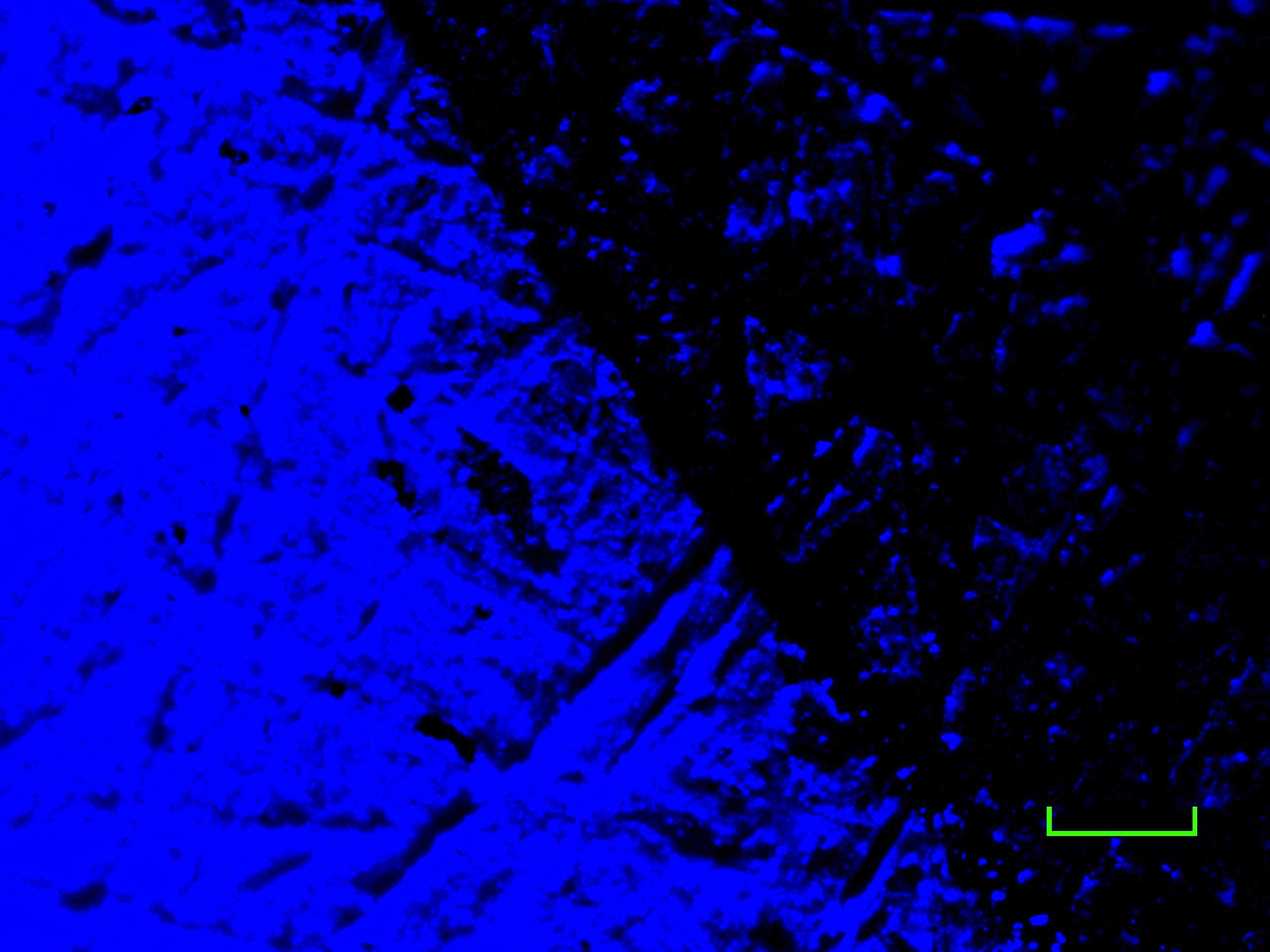}~\includegraphics[width=0.45\textwidth]{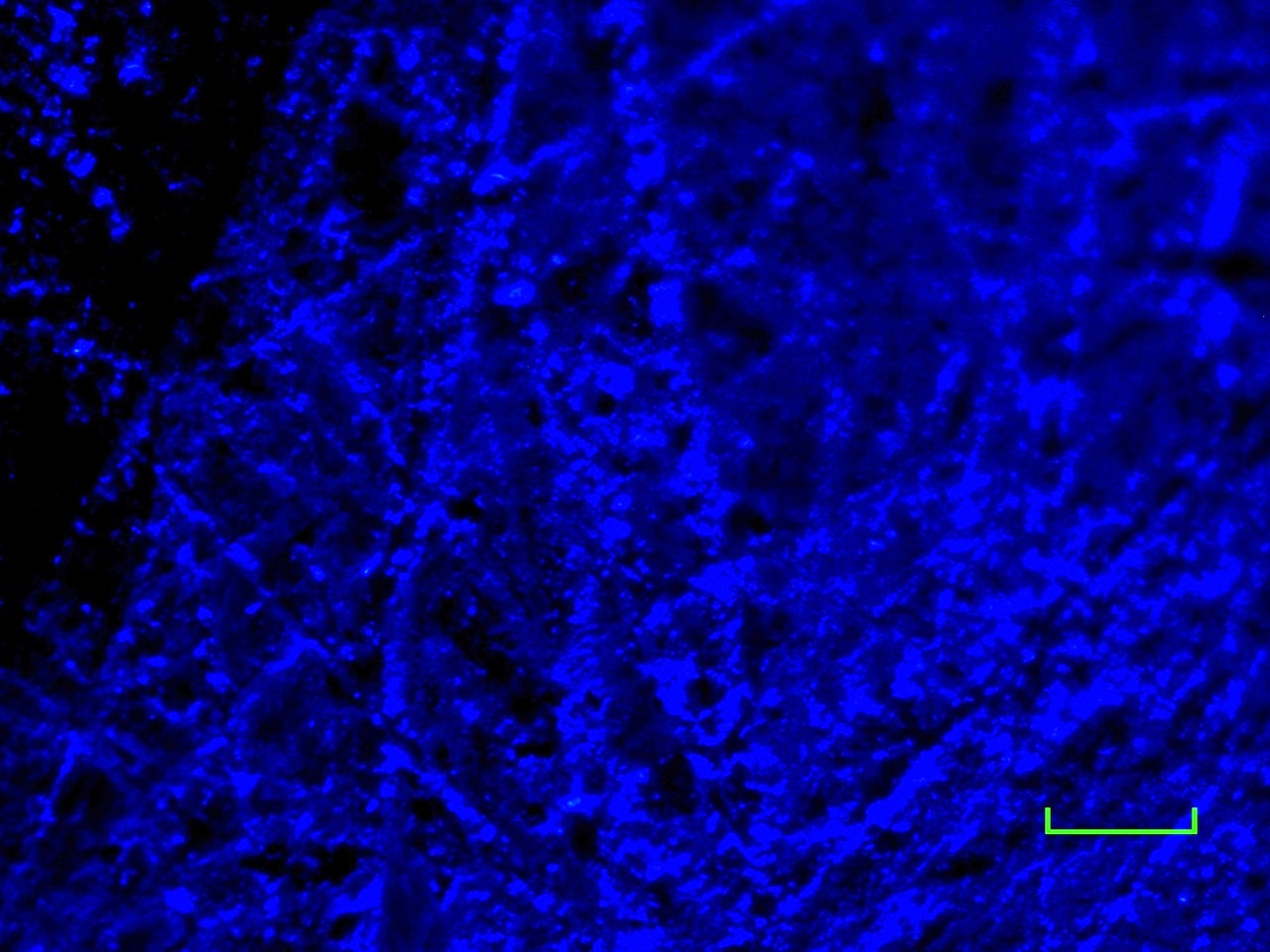}
\caption{Fluorescence microscopy image of the coating illuminated exclusively with a narrowband 355-nm light source. The photographs show in natural colour, wavelength shifted light, i.e. visible blue emission from TPB with any reflected UV component blocked with an additional filter. Black areas of the image correspond to uncoated acrylic, masked by the head of the bolt used to attach the witness sample to the stage. Good coverage of the substrate with TPB is shown, except for the masked areas. The scale bars corresponds to 61.5~$\upmu$m.}
\label{fig:fluorOptical1}
\end{center}
\end{figure}

\begin{figure}[h!]
\begin{center}
	\includegraphics[width=0.45\textwidth]{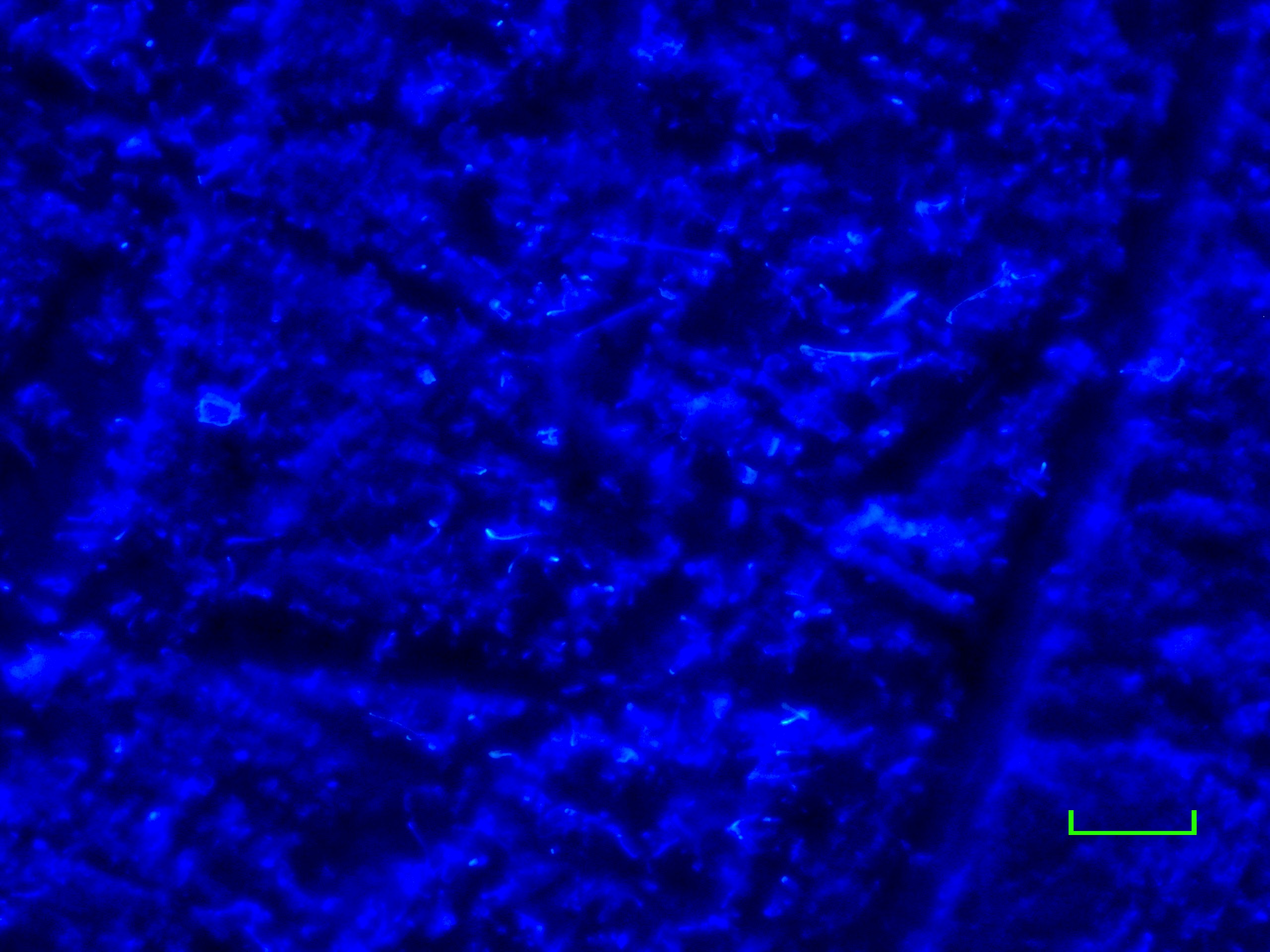}~\includegraphics[width=0.45\textwidth]{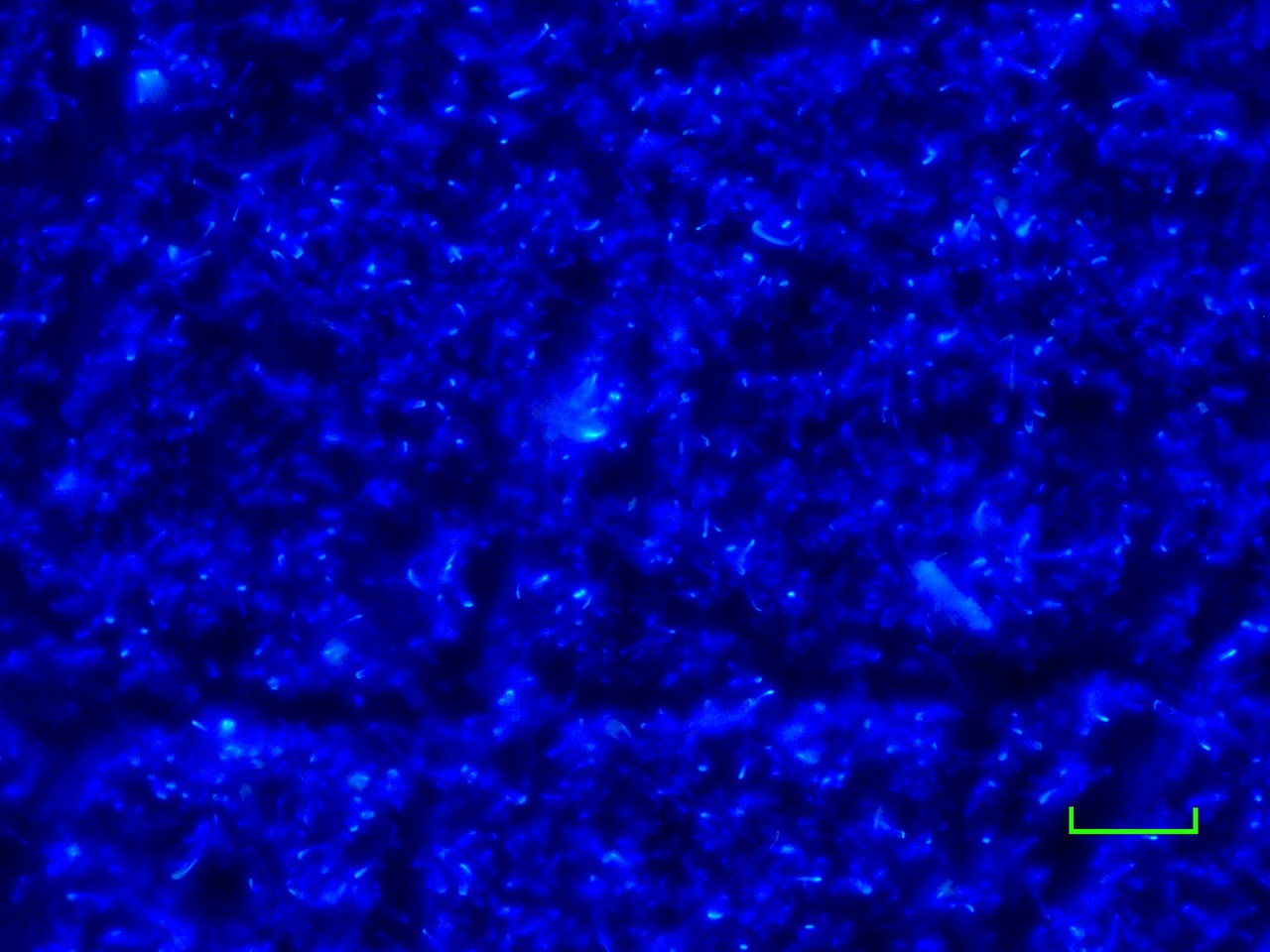}

\caption{Fluorescence microscopy image of the coating illuminated exclusively with a narrowband 355-nm light source. The photographs are zoomed in on central parts of the sample, to expose in finer detail more typical features of the coating and substrate. Crystalline needle-like $\alpha$-TPB structures of a few~$\upmu$m length are visible, as well as sparsely distributed larger crystals, also observed with other imaging techniques. The entire surface, including grooves from sanding, clearly visible on both images, is covered with a layer of TPB, which is evident from the blue glow (compare with uncoated acrylic shown in the previous figure). The scale bar corresponds to 31.25~$\upmu$m.}
\label{fig:fluorOptical2}
\end{center}
\end{figure}

Samples were also analysed with fluorescence microscopy, in which the coating was excited with light from an UVTOP355 365~nm UV LED filtered through a U-340 UV Bandpass Filter (Edmund Optics), creating a narrowband 355~nm source. Figure~\ref{fig:fluorOptical1} and~\ref{fig:fluorOptical2} show samples at two different magnifications. Re-emitted light from the TPB appears as blue, while areas left uncoated, such as in the upper right of the left panel in Figure~\ref{fig:fluorOptical1} which was masked from the TPB deposition by the mounting bolt for the sample, appear black. Some darker line features can be seen away from the masked area, but are the result of shadowing from the groove ridges during the imaging process. Fluorescence over the grooves, seen in magnification in Figure~\ref{fig:fluorOptical2} confirm coverage of TPB over the sanded acrylic surface features. 

In order to verify the mechanical stability of $\alpha$-TPB structures in cryogenic conditions, an additional liquid nitrogen dunk test was performed on a witness sample from the DEAP-3600 deposition. No adverse effects were observed; in direct comparison of optical microscopy images from a chosen region of the sample containing $\alpha$-TPB structures taken before and after exposure to liquid nitrogen, all of the abundantly present crystalline needles survived the test. Consistent with the results of earlier tests, no signs of macroscopic deterioration of the coating were observed.


\section{Summary}
\label{sec:Summary}
The DEAP-3600 experiment will perform a spin-independent dark matter search with a WIMP-nucleon cross section of 10$^{-46}$~cm$^{2}$ for a 100~GeV/$c^{2}$ WIMP mass. Due to the efficient absorption of VUV photons by most optical media, the wavelength shifter 1,1,4,4-tetraphenyl-1,3-butadiene was evaporatively deposited over the inner surface of the acrylic vessel to make visible the argon scintillation process. 

The technique presented for large-scale (approximately $9~\rm{m}^2$) thin-film deposition with adherence to thickness tolerances of several micrometers is, to best knowledge, novel, and significant development was undertaken to test the coating properties and overall deposition technique: cryogenic testing confirmed the robustness of the coating to exposure of cryogen given proper handling and preparation; small-scale limited-sampling tests confirmed the ability to achieve a controlled deposition to within a thickness non-uniformity tolerance of 20\% at larger-scales in multiple locations; large-scale testing confirmed the ability to coat in 4$\pi$ in a clean and non-detrimental manner for the underlying acrylic; imaging of sample slides identified surface morphology constraints and, in the final AV depositions, coverage at the targeted thickness of 3.00~$\pm$~0.02~$\upmu$m as estimated by the mass evaporated. DEAP-3600 is currently taking physics data, indicating a successful application of the TPB wavelength shifter.

\acknowledgments
This work is supported by the National Science and Engineering Research Council of Canada (NSERC), by the Canada Foundation for Innovation (CFI), and by the Ontario Ministry of Research and Innovation (MRI), and conducted on behalf of the DEAP collaboration, whose members' discussions and on-site involvement is greatly appreciated. We would like to thank Gregory Jerkiewicz for granting access to a surface profiler, Kevin Robbie and Chelsea Elliott for assistance with the SEM and optical imaging, Charles Cooney for assistance with optical imaging, Philippe Di Stefano for lending the UV light source for tests, Sanyasi Rao Bobbara for acquiring the AFM scans, and Neal Tennyson and Richard Bryan (Alfa Aesar) for fruitful discussions and adapting the TPB synthesis process to our requirements. We are grateful to SNOLAB and Vale Canada, Ltd. for excellent on-site support.

\end{document}